\documentclass[a4paper,aps,pra,notitlepage,superscriptaddress,longbibliography,nofootinbib,twocolumn,10pt]{revtex4-2}
\usepackage[utf8]{inputenc}
\usepackage{amsmath,amsthm,amssymb,amsfonts}
\usepackage{mathtools}
\usepackage{graphicx}
\usepackage{hyperref}
\usepackage{color}
\usepackage{float}
\usepackage{makecell}
\usepackage{enumitem}
\usepackage{mathdots}
\usepackage[dvipsnames]{xcolor}
\usepackage{hyperref}
\usepackage{braket}
\usepackage{bbm}
\usepackage{bm}
\usepackage[bb=boondox]{mathalfa}

\usepackage{tikz,tikz-3dplot,pgfplots}
\usetikzlibrary{shapes,arrows,patterns}
\usetikzlibrary{3d}
\usepackage{tikz-cd}

\newcommand{\Tr}{\operatorname{Tr}}

\newcommand{\ie}{{\it i.e.},\ }

\newcommand{\id}{\mathbb{1}} 

\newcommand{\im}{\operatorname{Im}}
\newcommand{\e}{\operatorname{e}}
\newcommand{\erf}{\operatorname{erf}}
\newcommand{\erfc}{\operatorname{erfc}}
\newcommand{\mi}{\mathrm{i}}
\newcommand{\sgn}{\operatorname{sgn}}

\pgfplotsset{compat=1.18}
\begin{document}
\title{Entanglement harvesting in quantum superposed spacetime}

\author{Anwesha Chakraborty}
\email{anwesha.chakraborty@unimelb.edu.au}
\affiliation{School of Mathematics and Statistics, The University of Melbourne, Parkville, VIC 3010, Australia}

\author{Lucas Hackl}
\email{lucas.hackl@unimelb.edu.au}
\affiliation{School of Mathematics and Statistics, The University of Melbourne, Parkville, VIC 3010, Australia}
\affiliation{School of Physics, The University of Melbourne, Parkville, VIC 3010, Australia}

\author{Magdalena Zych}
\email{magdalena.zych@fysik.su.se}
\affiliation{Department of Physics, Stockholm University, Roslagstullsbacken 21, 106 91 Stockholm, Sweden}

\begin{abstract}
We investigate the phenomenon of entanglement harvesting for a spacetime in quantum superposition, using two Unruh-DeWitt detectors interacting with a quantum scalar field where the spacetime background is modeled as a superposition of two quotient Minkowski spaces which are not related by diffeomorphisms. Our results demonstrate that the superposed nature of spacetime induces interference effects that can significantly enhance entanglement for both twisted and untwisted field. We compute the concurrence, which quantifies the harvested entanglement, as function of the energy gap of detectors and their separation.  We find that it reaches its maximum when we condition the final spacetime superposition state to match  the initial spacetime state. Notably, for the twisted field, the parameter region without entanglement exhibits a significant deviation from that observed in classical Minkowski space or a single quotient Minkowski space.
\end{abstract}

\maketitle

\section{Introduction}
Quantum mechanics exhibits subtle non-local characteristics, with quantum entanglement being the most well-known example. From the perspective of local observers, the vacuum state of any quantum field is inherently entangled, leading to correlated localized vacuum fluctuations. Studies have shown that correlations in the vacuum, as measured by local inertial observers, can, in principle, be strong enough to violate Bell-type inequalities~\cite{bell}. Additionally it has been demonstrated~\cite{entanglement_harvesting_menicucci,entanglement_harvesting_eduardo} that even when two uncorrelated local quantum systems, are placed in a quantum vacuum, they can become entangled by interacting with the entangled field---a phenomenon termed as `entanglement harvesting'. It has been a long posed question how non-locality in quantum mechanics depends on the global structure of space-time.

On the other hand, the equations of general relativity are inherently local, and therefore, they do not completely constraint the large-scale structure of the universe, such as topological properties. Various cosmological models lead to different global properties, including a diversity of spacetime topologies. It is conceivable that even a future theory of quantum gravity may not provide a definitive  description of the spatial topology of the universe.

The search for a consistent theory of quantum gravity has faced significant challenges, particularly due to the dynamic nature of spacetime, which complicates standard quantization methods. Given that gravity is inherently a theory of spacetime geometry, it stands to reason that any forthcoming theory should naturally incorporate the fundamental principles of quantum superposition in the fabric of spacetime. This integration should result in what is known as ``spacetime superpositions'', where different spacetimes, not connected by a global coordinate transformation, are combined in quantum superpositions. These investigations commonly focuses on the effects of spatial superpositions of massive objects~\cite{belenchia2018quantum,Bose,Marletto}, which involve semiclassical metrics that differ from an effective classical description of spacetime only by a coordinate transformation. Therefore, it is interesting to look at superpositions of spacetimes that are not diffeomorphic, meaning that the individual amplitudes represent independent solutions to Einstein’s field equations. While there is no complete quantum-gravitational theory for the emergence of such superpositions yet, we believe that it is reasonable to assume that such a theory should be able to describe the concept of a space-time superposition.  It was discussed in~\cite{Foo:2020jmi,Foo:2022ktn,Foo_BTZblackhole,Christodoulou_2019} that such spacetime are effectively equivalent to a single classical spacetime in which quantum systems are prepared and measured in appropriated quantum states. Recent studies~\cite{Giacomini:2021gei,Christodoulou_2019,Giacomini_2022,Henderson_2020,Foo_Udw_trajectory_superposition,Foo_BTZblackhole,PhysRevD.102.045002,Foo_super_Minkowski,Suryaatmadja,Goel} have investigated the quantum-gravitational phenomena arising from superpositions of mass or length parameters in periodically identified spacetimes, \ie spacetimes
characterized by distinct periodic boundary conditions. To explore these effects, they have coupled the relativistic quantum matter to quantum fields within such spacetime backgrounds, employing the Unruh-deWitt (UDW) particle detector model. Their findings reveal that the detector's response exhibits discrete resonances occurring at rational ratios of the superposed periodic parameters (length/mass). This observation sheds light on the intriguing interplay between quantum superposition, spacetime geometry, and the behavior of quantum matter, offering insights into the nature of spacetime at the quantum level.

In this paper, we investigate a pair of decoupled Unruh-deWitt (UDW) detectors within a background of superposed quotient Minkowski space.  A quotient Minkowski space has global nontrivial topological properties but is locally identical to a Minkowski space. In~\cite{Martin-Martinez}, it was demonstrated that two localized UDW detectors interacting with a quantum field can harvest entanglement, however the process is also influenced by the global structures of the underlying geometry. When compared to the entanglement harvested in simple Minkowski space, the results shown for the topologically nontrivial spaces showed to harvest more or less entanglement depending on the boundary conditions of the field (twisted vs. untwisted). Our goal in this manuscript is to  explore potential signatures indicating quantum superposition in the process of entanglement harvesting. For this, we apply the harvesting protocol from~\cite{Martin-Martinez} to the setting of two superposed Minkowski space, as introduced in~\cite{Foo_super_Minkowski}. We initialize the field theory state in the vacuum of free massless scalar field and then quantify the amount of entanglement in a pair of decoupled UDW detectors after interacting with the field and after conditioning on a chosen final spacetime state. Through this study, we demonstrate that the quantum superposition of geometry enhances the entanglement measured by the concurrence function when the detectors interact with either twisted or untwisted field. It is also shown that the concurrence will be maximal when the spacetime state measured along with the detector's state, is same as the initial spacetime state.\\

The paper is organized as follows: In section~\ref{sec:UDW-detectors}, we briefly review the conceptual idea of entanglement harvesting through two UDW detectors. In section~\ref{sec:quotient-space}, we then explain how to define the quotient Minkowski space, how to describe quantum fields living on it and how to encode superpositions of the field. In section~\ref{sec:unitary-evolution}, we then explain how to perturbatively compute the final entangled state of the two detectors after interacting with the quantum field under Gaussian switching function on a superposition of quotient Minkowski spaces. In section~\ref{sec:harvesting}, we then analyze the entanglement properties of this resulting state to determine how the parameters encoding our spacetime superposition affects entanglement harvesting. Finally, we summarize our findings in section~\ref{sec:discussion}.
\section{UDW Detectors and Entanglement Harvesting}\label{sec:UDW-detectors}
In this section, we review the interaction between two UDW detectors and a massless scalar quantum field in regular Minkowski spacetime. We then discuss the effect on entanglement harvesting when moving from regular Minkowski space to quotient Minkowski spaces with non-trivial topology.

\begin{figure}
    \centering
    \includegraphics[width=0.9\linewidth]{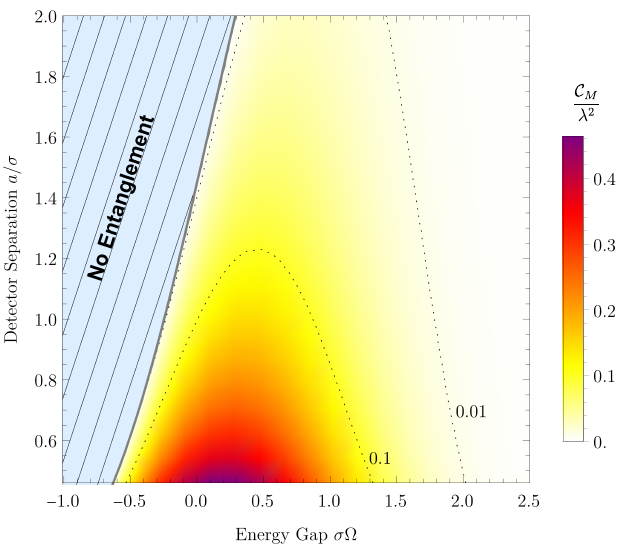}
    \caption{\textit{Density plot of concurrence in Minkowski space.} \\ $\mathcal{C}_M$ in Minkowski space is plotted against energy gap and detector separation, where no entanglement zone is separated (shaded area) from the entangled region and the contour lines of $\mathcal{C}_M$ are indicated at $0.1$ and $0.01$. The entanglement generation is maximum when detector separation as well as the energy gap is very small.}
    \label{fig:conc_minkowski}
\end{figure}

To understand the interaction between quantum field and the detectors, it is important to identify a measurement process that is both physically significant and mathematically straightforward. For this purpose, we consider a two-level atom as the measuring apparatus. Let $\ket{0}$ and $\ket{1}$ represent the ground and excited states of the atom (detector), separated by an energy gap $\Omega$. These states constitute an orthonormal basis for the detector Hilbert space $\mathcal{H}=\mathbb{C}^2$, associated with the atom's internal degrees of freedom. Its free evolution is governed by the Hamiltonian
\begin{equation}
    H_0=\frac{\Omega}{2}(\,\ket{1}\bra{1}-\ket{0}\bra{0}\,)\,.
\end{equation}
The interaction between a scalar field $\hat{\phi}(x(\tau))$ and the two level detector model is given by the Hamiltonian
\begin{equation}
H_{\mathrm{int}}=\lambda~\chi(\tau)~\big(\,\ket{1}\bra{0}+\ket{0}\bra{1}\,\big)\otimes\hat{\phi}(x(\tau))\,,\label{free-hamil}
\end{equation}
where $\lambda \ll 1$ is a weak coupling parameter that defines the strength of the interaction. $\chi(\tau) \in [0,1]$ is the switching function which signifies the duration of interaction between the atom and the scalar field. In the interaction picture, the interaction Hamiltonian is given by $H_I=\e^{\mi H_0 \tau}H_{\mathrm{int}}\e^{-\mi H_0 \tau}$. This is  the so called `Unruh-deWitt detector'~\cite{Pozas-Kerstjens:2016rsh,Louko_2008} model which encapsulates the majority of key aspects of light-matter interaction~\cite{PhysRevD.87.064038,PhysRevA.89.033835}. To study entanglement harvesting, we consider two such decoupled UDW detectors, labeled $A$ and $B$, associated with the detector Hilbert spaces $\mathcal{H}_A$ and $\mathcal{H}_B$ respectively, which follow the trajectories $x_A(\tau_A)$ and $x_B(\tau_B)$ and are parameterized by their proper times $\tau_A, \tau_B$. The detectors individually interact with a real massless scalar field $\hat{\phi}$ associated with the Hilbert space $\mathcal{H}_{\phi}$, through the interaction Hamiltonian
\begin{equation}
    H^{I}_D(\tau_D)= \lambda \chi_D (\tau_D) \Big(\sigma_+(\tau_D)+\sigma_{-}(\tau_D)\Big)\otimes \,\hat{\phi}(x_D(\tau_D)), \label{int-hamil}
\end{equation}
where $D=A,B$ and $\sigma_+(\tau_D)=\e^{\mi\, \Omega_D \tau_D} \ket{1}\bra{0}$, $\sigma_-(\tau_D)=\e^{-\mi\, \Omega_D \tau_D} \ket{0}\bra{1}$. Initially, we assume that both the detector and the scalar field system are in their respective ground states giving the total system's initial state as $\ket{\Psi_i}=\ket{0}_A\otimes \ket{0}_B\otimes\ket{0}_{\phi}$. The evolution of this detector-field system is described by the unitary operator $U=\hat{\mathcal{T}}\e^{-\mi\int\,dt\,[H_A^I(t)+H_B^I(t)]}$ where $H_A^I,H_B^I$ is defined in \eqref{int-hamil}. Using this time evolution operator, we can evolve the initial state $\ket{\Psi_i}$ to a final state $\ket{\Psi_f}$ and compute the reduced joint state of the detectors $A$ and $B$ as
\begin{equation}
     \rho_{AB}=\Tr_{\phi}\Big(U\ket{\Psi
    _i}\bra{\Psi_i}U^{\dagger}\Big)\,,\label{joint_state}
\end{equation}
where we trace over the scalar field Hilbert space. As shown in~\cite{entanglement_harvesting_menicucci,entanglement_harvesting_eduardo,Martin-Martinez}, this joint state is naturally represented as a $4\times4$ matrix with non-vanishing components in its diagonal and off-diagonal entries and thus often referred to as an `$X$'-state. To study entanglement harvesting, it suffices to expand this final state up to order $\lambda^4$ in our coupling parameter. It was further shown using the Peres-Horodecki criterion~\cite{PPT} that the joint state $\rho_{AB}$ is indeed entangled and only in the limit of large detector separation, it becomes a separable state $\rho_{AB}=\rho_A\otimes \rho_B$. Consequently, the concurrence $\mathcal{C}_M$ that quantifies the amount of entanglement does not vanish (see figure~\ref{fig:conc_minkowski}) and is maximal for small detector separation and and small energy gaps (relative to  $\sigma$). The key point is that the interaction between a quantum field and two initially unentangled UDW detectors gives rise to entanglement generation between the detectors. This entanglement arises due to the pre-existing entanglement in the field vacuum \cite{V_zquez_2014} and the exchange of quanta between the detectors during the interaction.

In the same paper~\cite{Martin-Martinez}, the importance of observing such effects in topologically non-trivial spaces was further discussed and entanglement harvesting was specifically calculated for two kinds of `cylindrical'~(spatially compactified) Minkowski space-time $M_0$ and $M_-$. For both cases, it was found that the transition probabilities of each detectors and also the concurrence deviate from the concurrence of Minkowski space, thus demonstrating that the global structure of space time affects the entanglement even if the detectors are interacting with the field locally and the local structure of space-time is identical to Minkowski. By evaluating the concurrence it was discovered that the entanglement generation increases and decreases for twisted and untwisted field respectively. The results demonstrated for $M_0$ cylindrical spaces,  mimics the results for field and detectors confined in cavity with appropriate boundary condition. Particularly it was analyzed
nonperturbatively in~\cite{PhysRevD.87.084062} and further in~\cite{PhysRevA.88.052310},
where it was shown that a combination of harvesting in
cavity complemented with communication yields a
sustainable source of quantum entanglement.\\

We now outline the setting of the present manuscript. As discussed in the introduction, the study of superposed space-time stands as a promising candidate for advancing our understanding of quantum gravity through a `bottom-up' approach. In the present work, we are interested to observe the `quantum' effects of superposition of space-time topology on entanglement generation from quantum field. Specifically, we shall take cylindrical Minkowski space $M_0$ with a compactified $z$ direction, in a superposed state of its characteristic length $L_1$ and $L_2$~\cite{Foo_super_Minkowski} as our background geometry and study the entanglement harvesting in two UDW detectors interacting with the massless scalar field following the approach of~\cite{Martin-Martinez}. In contrast to~\cite{Foo_Udw_trajectory_superposition}, where the state of \textit{single} UDW detector was considered, our focus lies on examining the effects of topological structures within a superposed background geometry on the entanglement harvesting in a \textit{pair of detectors}. Although our model is not fully realistic, it provides useful intuitive understanding and serves as a foundation for developing more realistic models involving curved geometries and provides promising direction to simulate effects of superposed geometries on quantum fields~\cite{Barcelo:2021nhs,Steinhauer_2016}.

\section{Quantum fields on quotient Minkowski space and its superposition}\label{sec:quotient-space}
The topologically nontrivial spacetime of our interest is the $M_0 \simeq M/J_{0}$ spacetime, formed by taking the quotient of standard Minkowski spacetime $M$ under the discrete isometry group $J_{0_L}: (t,x,y,z) \mapsto (t,x,y,z+L)$~\cite{Martin-Martinez, Langlois:2005nf} as below

 \begin{center}
   \begin{tikzpicture}
        \draw[thick,->,opacity=0.7] (0,0,0) -- (1.3,0,0) node[anchor=south east]{$x$}; 
    \draw[thick,->,opacity=0.7] (0,0,0) -- (0,1.2,0) node[anchor=north west]{$y$}; 
    \draw[thick,->,opacity=0.7] (0,0,0) -- (0,0,1.5) node[anchor=north]{$z$}; 
    \draw[thick,->,opacity=1.5](1.5,0.5)--(2,0.5);
     \begin{scope}[yshift=0mm,xshift=30mm]
        \coordinate (A1) at (0, 0, 0);
    \coordinate (B1) at (0.8, 0, 0);
    \coordinate (C1) at (0.8, 0.8, 0);
    \coordinate (D1) at (0, 0.8, 0);

    \coordinate (A2) at (0, 0, 1.3); 
    \coordinate (B2) at (1, 0, 1.3);
    \coordinate (C2) at (1, 1, 1.3);
    \coordinate (D2) at (0, 1, 1.3);
    \coordinate(L) at (1.2,0,0.7);

    \fill[blue!20, opacity=0.7] (A1) -- (B1) -- (C1) -- (D1) -- cycle;

    \fill[green!30, opacity=2.0] (A2) -- (B2) -- (C2) -- (D2) -- cycle;


    \draw[dashed] (B1) -- (B2);
    \draw[dashed] (C1) -- (C2);
    \draw[dashed] (D1) -- (D2);
\draw(L) node[scale=0.8]{$L$};
  \draw[decorate,decoration={brace,amplitude=3pt}] (B1)--(B2);
    \draw[thick,->,opacity=0.7] (0,0,0) -- (1.3,0,0) node[anchor=south east]{$x$}; 
    \draw[thick,->,opacity=0.7] (0,0,0) -- (0,1.2,0) node[anchor=north west]{$y$}; 
    \draw[thick,->,opacity=0.7] (0,0,0) -- (0,0,1.5) node[anchor=north]{$z$}; 
        \end{scope}
   \end{tikzpicture}
   \end{center}

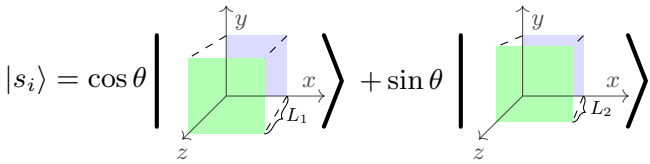
\begin{figure}[t!]
    \centering
    \begin{tikzpicture}
        \begin{scope}
        \draw (-.8,.01) node[scale=1.2,left]{$\ket{s_i}=\cos{\theta}$};
        \draw (0.4,0) node[scale=4]{$\ket{\hspace{4.8mm}}$};
        \begin{scope}[yshift=-2mm,xshift=1mm]
        \coordinate (A1) at (0, 0, 0);
    \coordinate (B1) at (0.8, 0, 0);
    \coordinate (C1) at (0.8, 0.8, 0);
    \coordinate (D1) at (0, 0.8, 0);

    \coordinate (A2) at (0, 0, 1.3); 
    \coordinate (B2) at (1, 0, 1.3);
    \coordinate (C2) at (1, 1, 1.3);
    \coordinate (D2) at (0, 1, 1.3);
\coordinate(L1) at (1.4,0.2,1.2);
    \fill[blue!20, opacity=0.7] (A1) -- (B1) -- (C1) -- (D1) -- cycle;

    \fill[green!30, opacity=2.5] (A2) -- (B2) -- (C2) -- (D2) -- cycle;

    \draw[dashed] (B1) -- (B2);
    \draw[dashed] (C1) -- (C2);
    \draw[dashed] (D1) -- (D2);
    \draw[decorate,decoration={brace,amplitude=3pt}] (B1)--(B2);
 \draw(L1) node[scale=0.8]{$L_1$};
    \draw[thin,->,opacity=0.7] (0,0,0) -- (1.3,0,0) node[anchor=south east]{$x$}; 
    \draw[thin,->,opacity=0.7] (0,0,0) -- (0,1.2,0) node[anchor=north west]{$y$}; 
    \draw[thin,->,opacity=0.7] (0,0,0) -- (0,0,1.5) node[anchor=north]{$z$}; 
        \end{scope}
        \end{scope}

        \begin{scope}[xshift=3.9cm]
        \draw (-.8,.01) node[scale=1.2,left]{$+\sin{\theta}$};
        \draw (0.5,0) node[scale=4]{$\ket{\hspace{4.8mm}}$};
        \begin{scope}[yshift=-2mm,xshift=1mm]
          \coordinate (A3) at (0, 0, 0);
    \coordinate (B3) at (0.8, 0, 0);
    \coordinate (C3) at (0.8, 0.8, 0);
    \coordinate (D3) at (0, 0.8, 0);
  \coordinate (A4) at (0, 0, 0.9); 
    \coordinate (B4) at (1, 0, 0.9);
    \coordinate (C4) at (1, 1, 0.9);
    \coordinate (D4) at (0, 1, 0.9);
    \coordinate(L2) at (1.4,0.2,0.95);
    \fill[blue!20, opacity=0.7] (A3) -- (B3) -- (C3) -- (D3) -- cycle;

    \fill[green!30, opacity=2.5] (A4) -- (B4) -- (C4) -- (D4) -- cycle;

    \draw[dashed] (B3) -- (B4);
    \draw[dashed] (C3) -- (C4);
    \draw[dashed] (D3) -- (D4);
      \draw[decorate,decoration={brace,amplitude=3pt}] (B3)--(B4);
 \draw(L2) node[scale=0.8]{$L_2$};
      \draw[thin,->,opacity=0.7] (0,0,0) -- (1.4,0,0) node[anchor=south east]{$x$};
    \draw[thin,->,opacity=0.7] (0,0,0) -- (0,1.2,0) node[anchor=north west]{$y$}; 
    \draw[thin,->,opacity=0.7] (0,0,0) -- (0,0,1.5) node[anchor=north]{$z$}; 
        \end{scope}
        \end{scope} 
    \end{tikzpicture}
    \caption{\textit{Initial state of a superposed spacetime.} The spacetime quantum state is initially an arbitrary superposition of $\ket{L_1}$ and $\ket{L_2}$ depending on the parameter $\theta$ where $L_1$ and $L_2$ characterize the compactification lengths of quotient Minkowski space.}
    \label{fig:spacetime-superposition}
\end{figure}

The action of $J_{0_L}$ allows us to identify spacetime points differing by a multiple of $L$ in the $z$ direction.  This spacetime has been explored in various contexts, including entanglement harvesting~\cite{Martin-Martinez}, symmetry-breaking in gauge theories~\cite{Toms:1983zb}, and the Casimir effect~\cite{PhysRevD.87.105012}. 
We consider the quantum field $\hat{\Phi}^L([x])$ on the quotient space (where $[x]\in M_0$ whereas $x\in M$)\footnote{Later, we will omit parentheses for the quotient space coordinate $[x]$, differentiation will be understood from the context of the automorphic field $\hat{\Phi}^L(x)$ and the usual field $\hat{\phi}(x)$.} constructed from the usual massless scalar field $\hat{\phi}(x)$ as the image sum given by~\cite{Banach:1979iy,Banach:1979wz}
\begin{equation}
    \hat{\Phi}^L([x])=\frac{1}{\sqrt{\mathcal{N}}}\sum_{n=-\infty}^{\infty} \gamma^n \hat{\phi} (J_{0_L}^n x)\,,\label{automorphic field}
\end{equation}
where $\mathcal{N}=\sum_n \gamma^{2n}$ is a normalization factor\footnote{Although the normalization $\mathcal{N}$ diverges, however this do not pose any practical obstacle. In all calculations, the normalization and the factors which is summed over various field modes and are formally also infinite ultimately cancel out, yielding finite, well-defined results, as explained in appendix B of~\cite{Alonso-Serrano:2021ydi}.} ensuring $[\hat{\Phi}(x),\dot{\hat{\Phi}}(x')]=\delta^3(x-x')+ \textrm{image terms}$\footnote{The calculation is underlined briefly as \begin{align*}
   [\hat{\Phi}(x),\dot{\hat{\Phi}}(x')]&=\frac{1}{\sum_n \gamma^{2n}}\sum_{m,n}\gamma^n\gamma^m[\hat{\phi}(J_{0_L}^n x),\dot{\hat{\phi}}(J_{0_L}^mx')]  \\
   &=\frac{1}{\sum_n \gamma^{2n}}\sum_{m,n} \gamma^n\gamma^{(n+m)}[\hat{\phi}(J_{0_L}^n x),\dot{\hat{\phi}}(J_{0_L}^nJ_{0_L}^mx')] \\
   &=\sum_{m}\gamma^{m}[\hat{\phi}(x),\dot{\hat{\phi}}(J_{0_L}^mx')]=i\delta^3(x-x')+\text{image terms}
\end{align*}}
 and $\gamma$ is representation of the discrete group $J_0$ taking values in $SL(\mathbb{R})=\{1,-1\}$  denoting untwisted and twisted field respectively. Untwisted fields satisfy standard periodic boundary conditions and remain invariant under the action of the symmetry group, while twisted fields acquire a phase when subjected to the symmetry such as
\begin{align}
    \hat{\phi}^L(z+nL)&=\hat{\phi}(z)&& \textrm{for untwisted fields},\nonumber\\
    \hat{\phi}^L(z+nL)&=(-1)^n\,\hat{\phi}(z) &&\textrm{for twisted fields.}\label{twisted field}
\end{align}
As our goal here is to explore the effect of spacetime superposition on entanglement harvesting and we follow~\cite{Foo_super_Minkowski} to construct the two dimensional `spacetime Hilbert space' $\mathcal{H}_S=\mathrm{span}\{\ket{L_1},\ket{L_2}\}$ where arbitrary superpositions (as figure-\ref{fig:spacetime-superposition}) of $\ket{L_1}$ and $\ket{L_2}$ can be taken to represent a quantum state of $M_0$ spacetime.

Treating spacetime quantum superpositions in this framework can be understood as an effective description of quantum spacetime without the requirement of having a complete theory of quantum gravity, where we combine the standard description of quantum field theory in curved spacetime with the ability to encode simple quantum superpositions of certain spacetimes. While this is certainly far away from a complete quantum gravitational description and we thus do not expect to capture all arising phenomena, it is expected that spacetime superposition is at least one of the features in quantum gravity~\cite{Foo_BTZblackhole,Foo_super_Minkowski,Foo:2020jmi} and so our framework provides a proof of concept to explore this effect quantitatively. While we focus on the case where our spacetime state is in a superposition of just \emph{two} quotient Minkowski spaces, its generalization to more complicated superpositions is straightforward.

Let us emphasize here that~\eqref{automorphic field} provides an embedding of field operators in quotient Minkowski space into the operator space of regular Minkowski space. This is important when considering superpositions, because it means that we do not need to consider different field theory Hilbert spaces associated to the different compactification lengths $L_i$. Instead, there is a single field theory Hilbert space $\mathcal{H}_\phi$, but depending on the compactification length $L_i$ we need to construct our field operator $\hat{\Phi}(x)$ at $x$ according to~\eqref{automorphic field}. On the joint Hilbert space $\mathcal{H}_\phi\otimes\mathcal{H}_{S}$, we need to condition our field operator on the state of the spacetime, yielding the relation
\begin{align}
    \hat{\Phi}(x)=\sum_i \hat{\Phi}^{L_i}(x)\otimes \ket{L_i}\bra{L_i}\,,
\end{align}
where $\hat{\Phi}(x)$ represents field operators on the joint field-spacetime Hilbert space $\mathcal{H}_\phi\otimes\mathcal{H}_{S}$, which automatically conditions $\hat{\Phi}$ to be $\hat{\Phi}^L$ for spacetime states $\ket{L}$. 
\section{Joint state of detectors under unitary evolution}\label{sec:unitary-evolution}
We now study the unitary evolution of two decoupled UDW detectors interacting with a massless scalar field in the superprosed Minkowski background. 
The full Hilbert space of the system involving the detectors, fields and space-time states is given by $\mathcal{H}_A\otimes\mathcal{H}_B\otimes\mathcal{H}_{\phi}\otimes\mathcal{H}_S$.
The system is described the following interaction Hamiltonian 
\begin{align}
    H^{I}_D(\tau_D)&= \lambda \chi_D (\tau_D) \Big(\sigma_+(\tau_D)+\sigma_{-}(\tau_D)\Big)\nonumber\\
    &\hspace{1.5cm}\otimes \sum_{i=1,2}\, \hat{\Phi}^{L_i}(x_D(\tau_D))\otimes \ket{L_i}\bra{L_i},\label{interaction Hamiltonian}
\end{align}
with $D= A,B$ and where the elements in the Hamiltonian has been described earlier in~\eqref{free-hamil} and~\eqref{int-hamil}. More details on realization of such thing is described in~\cite{Foo_super_Minkowski}.
The total interaction Hamiltonian for the two detector-field-space-time system is given by
\begin{equation}
    H_{\mathrm{tot}}^{I}= H_{A}^{I}(\tau_A)\otimes \id_B+\id_A\otimes  H_{B}^{I}(\tau_B).
\end{equation}
We denote $\id_A$ and $\id_B$ as the identity operators acting on the Hilbert spaces $\mathcal{H}_A$ and $\mathcal{H}_B$ respectively. Henceforth, we will omit the indices and refer to them based on their positions. Additionally, we will simplify notation by omitting `I' in the superscript of the Hamiltonian to reduce index cluttering.

The initial state of the full system is taken to be
\begin{align}\ket{\psi_i}&=\ket{0,0}\otimes\ket{0}_F\otimes\ket{s_i},\label{initial state}
\end{align}
where $\ket{0,0} \equiv \ket{0}_A\otimes \ket{0}_B,~\ket{0}_F$ is the Minkowski vacuum and
\begin{equation}
\ket{s_i}=\cos\theta\ket{L_1} + \sin\theta\ket{L_2}\label{s_i},
\end{equation}is an arbitrary superposition (see figure-\ref{fig:spacetime-superposition}) of the two space-time states. Note that we choose global Minkowski vacuum corresponding both the fields $\hat{\Phi}^{L_1}$ and $\hat{\Phi}^{L_2}$. The two sets of modes share the same vacuum state because the mode functions in each spacetime differ only by an overall phase factor.

\begin{figure}[t]
    \centering
\begin{tikzpicture}

    \def\distance{1.5}
    \def\planeSize{2.5}
    \coordinate(A) at (0,0);
\coordinate(B) at (0.9,0.8);
    \fill[blue!20, opacity=0.7] (0,0) -- ++(2,-1) -- ++(\planeSize,0) -- ++(-2,1) -- cycle;
    \fill[green!30, opacity=0.7] (0,\distance) -- ++(2,-1) -- ++(\planeSize,0) -- ++(-2,1) -- cycle;
    \draw (0,1.5) node[left]{$L$};
\draw[thick,red](A) -- (B);
\draw(0.5,0.28) node[right]{$~\alpha$};
\draw[thick,->,black] (.7,0) arc (0:40:7mm);
    \draw[->] (0,0) -- (0,2) node[right] {$z$};
    \draw[->] (0,0) -- (3,0) node[above] {$x$};
    \draw[->] (0,0) -- (2.5,-1.2) node[right] {$-y$};
 \fill[blue!80] (0,0) circle [radius=0.08cm] node[left]{\tiny{$~A$}};
 \fill[blue!80] (0.9,0.8) circle [radius=0.08cm] node[right]{\tiny{$B$}};
 \draw[](0.38,0.65) node{$a$};
\end{tikzpicture}
    \caption{\textit{Detector orientation with respect to compactified direction $z$.} Without loss of generality we can take detector $A$ at spatial origin $\Vec{x}_A=(0,0,0)$ and parametrize the coordinates of the detector $B$ as $\Vec{x}_B=(a\cos\alpha,0,a\sin\alpha)$, where $\alpha$ denotes the angle between the line connecting two detectors and $x$-axis.}
    \label{fig:detector_angle}
\end{figure}
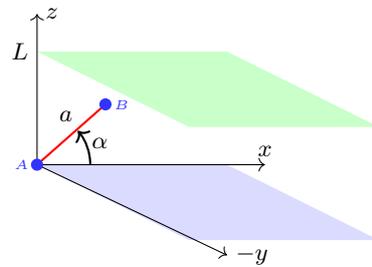

The unitary operator responsible for evolution of the initial quantum state in the interaction picture is given by the following
\begin{align}
    U&= \hat{\mathcal{T}}\Bigg[\exp\Big\{-\mi\int \,\,dt\,\,\Big(\frac{d\tau_A}{dt}\Big)\,H_A(\tau_A(t)) \otimes \id\nonumber\\
    &\qquad\qquad\qquad+\id\otimes \Big(\frac{d\tau_B}{dt}\Big)\,H_B(\tau_B(t)) \Big\}\Bigg]\label{unitary}
\end{align}
with $D= A,B$ and where we have chosen to evolve the field and detectors. After time evolution of \eqref{initial state} under \eqref{unitary}, the final state is be given by
\begin{equation}
    \ket{\psi_f} =\sum_n \lambda^n \ket{\psi_f^{(n)}} =\sum_{n} U_n \ket{\psi_i}\,, \label{finalstate}
\end{equation}
where the $U_n$s are the terms of order $\lambda^n$ in the unitary operator $U$. We can find the reduced density matrix (appendix~\ref{app_density matrix}) for the joint detector state by tracing out the field degrees of freedom and by conditioning on the control degree of freedom of space-time states \ie 
\begin{equation}
\ket{s_f}=\cos\phi\ket{L_1}+\sin\phi\ket{L_2}.\label{eq:s_f}
\end{equation}The joint state of the detectors $\rho_{AB}=\Tr_{\Phi}[\bra{s_f}U\ket{\psi_i}\bra{\psi_i} U^{\dagger}\ket{s_f}]$ written in the basis $\ket{0,0}, \ket{0,1}, \ket{1,0}, \ket{1,1}$ and up to $\lambda^2$ order (see appendix~\ref{app_density matrix} for details) is given by 
\begin{equation}
    \rho_{AB}= \begin{pmatrix}
        P^G&0&0&X\\
        0&P^{E}_{B}&C&0\\
        0&C^*&P^{E}_{A}&0\\
        X^*&0&0&0
    \end{pmatrix}\label{density matrix}
\end{equation}
where the respective terms are given by
\small
\begin{align}
    &P^G=(a+b)^2-\lambda^2\Big[(a^2+ab)\sum_{D}P_D^{L_1}+(b^2+ab)\sum_{D}P_D^{L_2}\Big]\label{pground}\\&P^{E}_{D}=\lambda^2(a^2~P_D^{L_1}+b^2~P_D^{L_2}+2a b~P_D^{L_1L_2}),\, \text{where}\label{total transprob}\\
   &P_D^{L_i}= \int dt\,dt'\nu_D(t)\bar{\nu}_D(t')\,W^{L_i}\Big(x_D(t),x_D(t')\Big),\,\,\text{and}\label{transprob}\\ 
  & P^{L_1L_2}_D=\int dt\,dt' \nu_D(t)\bar{\nu}_D(t')\,W^{L_1L_2}\Big(x_D(t'),x_D(t)\Big),\label{interference tranprob}\\
   &C= \lambda^2\int dt\,dt' \nu_A(t)\bar{\nu}_B(t')\Big[a^2~W^{L_1}\Big(x_A(t),x_B(t')\Big)\nonumber\\
   &+b^2 ~W^{L_2}\Big(x_A(t),x_B(t')\Big)+2ab  ~W^{L_1L_2}\Big(x_A(t),x_B(t')\Big)\Big]\label{c term}\\
   &X=-2\lambda^2\int dt\,dt'\nu_A(t)\nu_B(t')\Big[(a^2+ab)W^{L_1}(x_B(t'),x_A(t))\nonumber\\
   &\hspace{80pt}+(b^2+ab)~W^{L_2}(x_B(t'),x_A(t))\Big]\label{x term}
\end{align}
\normalsize
with $\nu_D(t)=\chi_D(t)\e^{-\mi \Omega_D\tau_D(t)}$ and $a=\cos\theta\cos\phi,\, b=\sin\theta\sin\phi$. The Wightman's function or the two-point correlators $W^{L_i}(x_D(t),x_{D'}(t'))$ in the quotient Minkowski space $M_0$ with periodicity $L_i$ in the $z$-direction is explicitly given by~\cite{Foo_super_Minkowski},
\begin{align}
    &W^{L_i}(x_D(t),x_{D'}(t'))=\bra{0}\hat{\Phi}^{L_i}(x_D)\hat{\Phi}^{L_i}(x'_{D'})\ket{0}\nonumber\\
    &\hspace{3mm}=\frac{1}{\mathcal{N}}\sum_{n,m}\gamma^n\gamma^m \bra{0}\,\hat{\phi}(J^n_{0_{L_i}}x_D)\,\hat{\phi}(J^m_{0_{L_i}}x'_{D'})\,\ket{0},\label{wightmansingle}
\end{align}
and $W^{L_1L_2}(x_D(t),x_{D'}(t'))$ is the two point correlation  between two fields, $\hat{\Phi}^{L_1}(x_D)$ and $\hat{\Phi}^{L_2} (x'_{D'})$, which
are individually quantized on the spacetime backgrounds in superposition and is given by
\begin{align}
&W^{L_1L_2}(x_D(t),x_{D'}(t'))=\bra{0}\hat{\Phi}^{L_1}(x_D)\hat{\Phi}^{L_2}(x'_{D'})\ket{0}\nonumber\\
&\hspace{0.6cm}=\frac{1}{\mathcal{N}}\sum_{n,m}\gamma^n\gamma^m \bra{0}\,\hat{\phi}(J^n_{0_{L_1}}x_D)\,\hat{\phi}(J^m_{0_{L_2}}x'_{D'})\ket{0}.\label{wightmandouble}
\end{align}
Using~\eqref{wightmansingle} and~\eqref{wightmandouble}, we can evaluate the matrix components $P^E_D,C,X$ of \eqref{density matrix} and for that we consider two identical detectors with $\Omega_A=\Omega_B,\,\chi_A(t)=\chi_B(t)$ separated by a spatial distance $a$ (see figure~\ref{fig:detector_angle}).
With that the matrix components up to order $\lambda^2$ are calculated in appendix~\ref{app_density matrix}.

Note that~\eqref{density matrix} is not a normalized density matrix ($\Tr(\rho_{AB})\ne 1$) as we are restricting ourselves to the subset of outcomes conditioned on the superposed state $\ket{s_f}$. If on the other hand, we trace over the control system, rather than performing the measurement, then it  would give $\Tr(\rho_{AB})=1$. To have a normalized density matrix one must divide it with $\Tr (\rho_{AB})$ (see appendix~\ref{normalization}). However, even after normalization $\rho_{AB}$ still does not satisfy positivity condition of density matrix. To solve the situation we need to consider the density matrix up to order of at least $\lambda^4$ and by doing so we land up with the following density operator
\begin{equation}
     \rho_{AB}= \begin{pmatrix}
        P^G+E&0&0&X\\
        0&P^{E}_B-E&C&0\\
        0&C^*&P^{E}_A-E&0\\
        X^*&0&0&E
        \end{pmatrix}\,,\label{xstate}
\end{equation}
where $E$ is given by (see~\cite{Martin-Martinez} for details)
\begin{equation}
    E=P_A^EP_B^E+\vert C\vert ^2+\vert X\vert ^2.\label{form E}
\end{equation}

\begin{figure}[t]
    \centering
    \begin{tikzpicture}
        \draw (0,0) node{\includegraphics[width=1.1\linewidth]{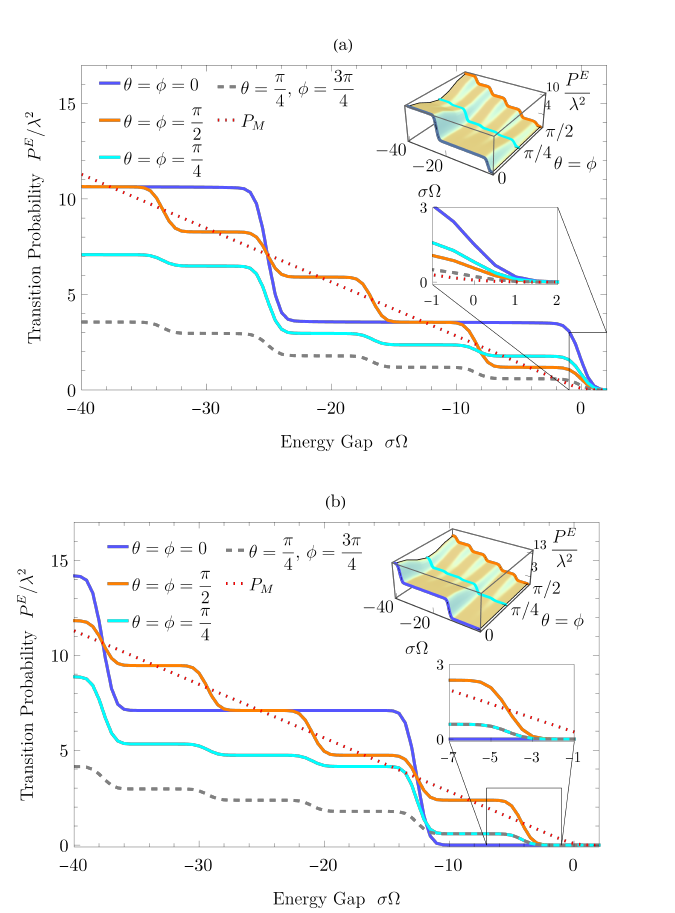}};
    \end{tikzpicture}
    \caption{\textit{Transition probability  $P^E$~\eqref{total transprob} in a superposed  and $P_M$ in regular Minkowski space.} Transition probability of a detector interacting with (a) untwisted field and (b) twisted field is plotted  as a function of energy gap $\Omega$ (in units of $\sigma$) for superposed space with $L_1=\frac{\sigma}{4}$, $L_2=\frac{3\sigma}{4}$. The angles $(\theta,\phi)$ parametrize the initial and the final state of the spacetime superposition, respectively, as shown in~\eqref{s_i} and~\eqref{eq:s_f}. We are mostly interested in the case $\theta=\phi$, where we see transition probability to vary smoothly between the states $\ket{L_1}$ for $\theta=\phi=0$ and $\ket{L_2}$ for $\theta=\phi=\pi/2$ (see inset). However, we also illustrate the case (dashed line), where we start with $\theta=\pi/4$ and then condition the measurement on finding the state with $\phi=3\pi/4$.}
    \label{fig:probability-E}
\end{figure}

The matrix in \eqref{xstate} is the anticipated `X-state'~\cite{Pratt_2004,Wang_2006,Xstate1}, the name being due to the density matrix’s resemblance with the letter ‘X’. which is normalized and satisfies positivity condition \footnote{The components $P^E_D,C,X$ of $\rho_{AB}$ in \eqref{xstate} get some more contributions of order $\mathcal{O}(\lambda^4)$, but we neglect those terms as they are of negligible value as $\lambda\ll 1$ and do not affect the positivity condition.} and was also encountered in~\cite{Martin-Martinez}. Also it can be checked that by tracing out the individual detector's Hilbert space one can find the reduced state of detector $A$ or $B$ as
\begin{equation}
    \rho_A =\Tr_B(\rho_{AB})=\begin{pmatrix}
        1-P_A^E&0\\0&P_A^E 
    \end{pmatrix}.
\end{equation}
and vice versa. At this point one can check that this density matrix matches with the one in~\cite{Foo_super_Minkowski} where they had considered a single detector in superposed Minkowski background. Next, we discuss some special cases as shown in the figure~\ref{fig:probability-E}:

(i) By setting $ \theta = \phi = 0 $, we retrieve the transition probability for a detector $ P_D^{L_1} $ in a single cylindrical Minkowski space with compactification length $ L_1 $.
   
(ii) Similarly, for $ \theta = \phi = \frac{\pi}{2} $, we obtain the transition probability $ P_D^{L_2} $.

(iii) Setting $\theta = \phi = \frac{\pi}{4} $ prepares the initial spacetime state and final measurement control state as $ \ket{s_i} = \ket{s_f} \to \ket{+}= \frac{1}{\sqrt{2}} (\ket{L_1} + \ket{L_2}) $, a symmetric superposition of $ \ket{L_1} $ and $ \ket{L_2} $, yielding the transition probability $ P^E_+ = \frac{1}{4}(P^{L_1} + P^{L_2} + 2P^{L_1L_2}) $ as calculated in~\cite{Foo_super_Minkowski}.

(iv) Setting $ \theta = \frac{\pi}{4} $ and $\phi = \frac{3\pi}{4} $ prepares the initial spacetime state in $\ket{+}$, while the final spacetime state is an anti-symmetric superposition, $\ket{s_f} \to\ket{-}= \frac{1}{\sqrt{2}} (\ket{L_1} - \ket{L_2})$. This yields the transition probability $ P^E_- = \frac{1}{4}(P^{L_1} + P^{L_2} - 2P^{L_1L_2})$ as shown in~\cite{Foo_super_Minkowski}.

There are some pertinent observations from figure~\ref{fig:probability-E} as follows: As the compactification length \( L \) increases, the transition probability becomes more oscillatory, and eventually for very large $L$, it mimics the Minkowski transition probability \( P_M \) (red dotted line) giving the \( L \to \infty \) limit of the transition probability in~\eqref{trans prob final}. Also the transition probability decreases and asymptotically approaches zero with energy gap $\sigma\Omega$ near origin, much earlier for twisted fields compared to untwisted fields. This early drop in probability for twisted fields reflects the reduced correlation strength in the field, which results from altered boundary conditions and interference effects. The Wightman function \( W(x,x') \), which encodes field correlations and governs detector interactions, is modified in twisted fields due to the phase factor \( (-1)^n \) (as seen in~\eqref{twisted field}). This phase induces destructive interference, reducing the detector's transition probability, which directly depends on \( W(x,x') \). That is why, near the origin of \( \sigma \Omega\) axis, the probability for untwisted fields in symmetrically superposed space (cyan) is notably higher than that of the Minkowski space, while the opposite is true for twisted fields. This difference explains the larger entanglement region for twisted fields in the superposed space compared to untwisted fields as well as regular quantum field in Minkowski space, as shown and discussed in more details later in context of figure~\ref{fig:concurrence_density}.

\section{Logarithmic negativity and entanglement harvesting}\label{sec:harvesting}
When a pair of UDW detectors interacts with the quantum field, the detector pair and the field are initially in a product state~\eqref{initial state}. Since there is no direct interaction between the detectors, any entanglement that arises between the detectors originates from the entanglement already present in the vacuum state of the field. We now examine how this entanglement generation is influenced by the superposed structure of spacetime. Let us note that the components $C$ and $X$ in~\eqref{xstate}, derived in~\eqref{x final} and~\eqref{c final} in appendix~\ref{app_density matrix}, depend on the spatial distance between the two detectors $a$ and the compactification lengths of the space-times $L_i$s. They vanish, \ie $C,X \to 0$, if either of these lengths $a$ or $L_i$ becomes very large. In this limit, we can write the joint density operator as tensor product of the individual density operators, \ie $\rho_{AB}=\rho_A\otimes\rho_B$, and there is no entanglement between the detectors.
\begin{figure}[t]
 \begin{center}
    \includegraphics[width=1\linewidth]{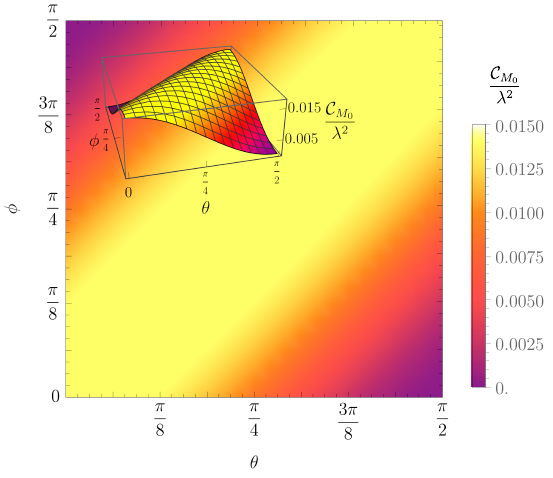}
    \caption{\textit{Dependence of concurrence on varying initial and final spacetime states.} $\mathcal{C}_{M_0}$ for superposed space is plotted against $\theta$ and $\phi$ showing that it is always maximum around $\theta=\phi$ meaning the final spacetime state on which the joint detector's state is conditioned, is same as initial spacetime state \ie $\ket{s_i}=\ket{s_f}$ where used parameters are $L_1/\sigma=3.8, L_2/\sigma=4, a/\sigma=1.6$ and $\sigma\Omega=1.4$. The behavior is here shown only for untwisted field as it follows the same nature for twisted field.}
    \label{fig:conc_angle}
     \end{center}
\end{figure}
\begin{figure}[t]
 \begin{center}
    \includegraphics[width=1\linewidth]{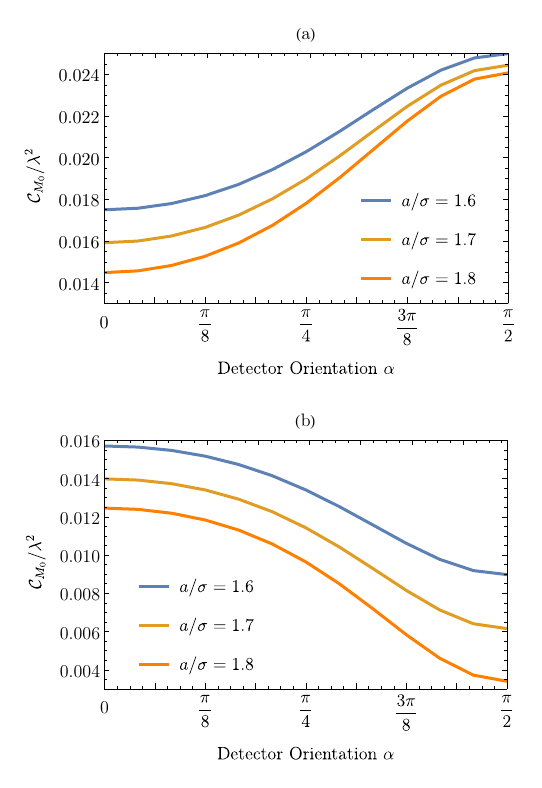}
    \caption{\textit{Dependence of concurrence with varying detector pair orientation.} In superposed space with  with $\theta = \phi = \pi/4$ and $L_1/\sigma = 3.8$, $L_2/\sigma = 4$, $\mathcal{C}_{M_0}$  is plotted as a function of $\alpha$ for different values of detector separation $a/\sigma$. (a) For untwisted field the concurrence increases as the line connecting the two detectors aligns more closely with the compactified $z$ axis (see figure~\ref{fig:detector_angle}), reaching its maximum when the alignment is parallel to the compactified direction, (b) For twisted field it follows the exact opposite nature giving the lowest value of concurrence when the detector pair is parallel to the compactified direction.}
    \label{fig:conc_detector separation}
     \end{center}
\end{figure}

To examine how much entanglement the detectors are able to harvest, we must quantify it, for which we will use the negativity or concurrence. According to the Peres-Horodecki criterion, a joint state is entangled if and only if its partially transposed matrix with respect to any of the contributing states has at least one negative eigenvalue and the amount of entanglement can be measured by the so-called \emph{negativity} $\textbf{N}$, computed as
\begin{equation}
    \hspace{-3mm}\textbf{N}=\frac{\Vert \rho^{\Gamma_A}\Vert -1}{2}=\sum \textrm{(negative eigenvalues of}~\rho^{\Gamma_A})\,,
\end{equation}
where $\rho^{\Gamma_A}$ is the partial transpose of $\rho_{AB}$ with respect to the subsystem $A$. For two identical detectors, the Peres-Horodecki criterion gives rise to the conditions (up to $\mathcal{O}(\lambda^2)$)
\begin{equation}
    |X| > P^E,\qquad |C| >\sqrt{E}\label{conditions}].
\end{equation}
\begin{figure*}[!t]
     \centering
        \includegraphics[width=0.9\linewidth]{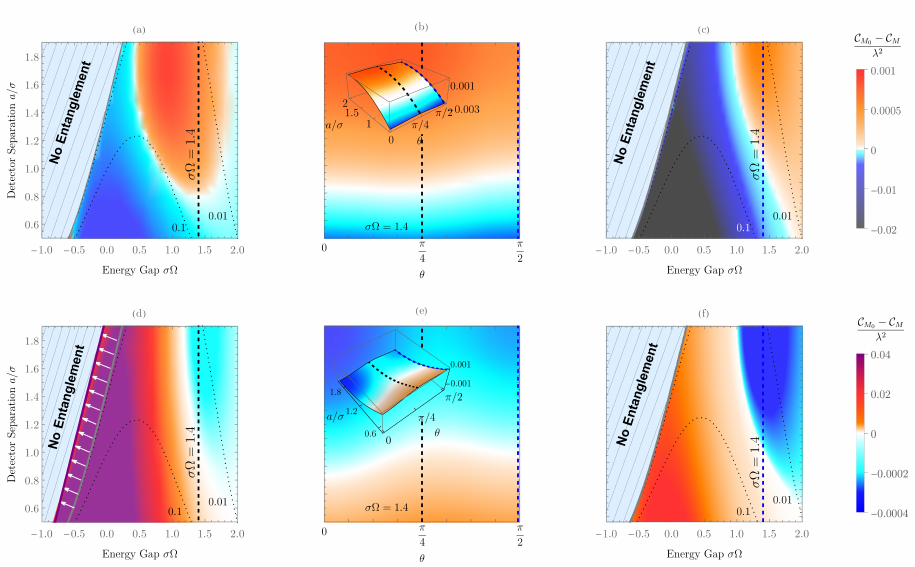}
        \caption{\textit{Density plot of Concurrence with detector separation and energy gap.} The difference in concurrence between the final state of two detectors in cylindrical (single and superposed) spacetime $M_0$ and in Minkowski space $M$, denoted as $\mathcal{C}_{M_0} - \mathcal{C}_M$, is plotted as a function of the detector separation $a/\sigma$ and energy gap $\sigma\Omega$.
 (a) and (d) shows the cases when detectors are coupled to an untwisted and twisted field respectively in superposed Minkowski space with $L_1/\sigma = 3.8$ and $L_2/\sigma = 4$, while  (c) and (f) shows the same for a single cylindrical space with $L/\sigma = 4$. The dotted contour lines shown in (a,c,d,f) are the contour lines of concurrence in Minkowski space $\mathcal{C}_M$ shown previously in figure~\ref{fig:conc_minkowski} which serve as a benchmark. In (b) and (e) we have chosen a particular value of $\sigma\Omega=1.4$ for which the difference in concurrence is varied with the  detector separation (along $y$ axis) and $\theta$ (along $x$ axis), where the values $\theta=\pi/4$ and $\theta=\pi/2$ indicates superposed and single quotient space respectively as two limiting cases pictorially shown in left and right of each plot for untwisted and twisted fields. For all the plots above we used $\alpha=0$. Note that entanglement region is enlarged in case of twisted field in superposed background with $\theta=\pi/4$.}
        \label{fig:concurrence_density}
\end{figure*}%
By inspecting~\eqref{form E}, it is evident that the second condition is never satisfied. So by satisfying the first condition in~\eqref{conditions}, we will be able to tell whether our system is entangled or not. A measure of entanglement is given by concurrence, which provides a quantitative means to assess the degree of entanglement between two qubits. For two identical detectors  ($\Omega_A=\Omega_B,\,\chi_A=\chi_B$) concurrence is quantified in terms of negativity as $\mathcal{C}=2\textbf{N}$ which in this case is given by
 \begin{equation}
     \mathcal{C}= 2\textbf{N}= \textrm{Max}\Big(0,2(|X|-P^E)\Big)\,,
 \end{equation} 
meaning we will only have nonzero concurrence and thus observe entanglement when the first condition of~\eqref{conditions} is satisfied. So we compare the concurrence $\mathcal{C}(\rho_{AB})$ in the final state $\rho_{AB}$ of two static detectors in superposed spacetime with that of single cylindrical spacetime and Minkowski spacetime. We start by examining the nature of concurrence in figure~\ref{fig:conc_angle} with varying $\theta$ and $\phi$, which are the angles quantifying the parameters of the initial and final (measuring) quantum states of the spacetime. We thereby demonstrate that we could attain maximal entanglement around in the region where $\theta=\phi$. In the following parts, we will therefore focus on superposed spacetimes with $\theta=\phi$ and specifically two cases: $\theta=\phi=\pi/4$ (symmetric superposition of $\ket{L_1}$ and $\ket{L_2}$) and $\theta=\phi=\pi/2$ (giving single quotient space with cylindrical length parameter $L/\sigma=4$).

In order to understand the effect of the spacetime superposition state \(\ket{s_i}=\ket{s_f}\) on the amount of harvested entanglement, we compute the concurrence as a function of the superposition angle \(\phi=\theta\) in figure~\ref{fig:conc_detector separation} and observe for an untwisted field (figure~\ref{fig:conc_detector separation}(a)) that entanglement increases as the detector pair aligns with the compactified direction (\ie $\alpha = \pi/2$ in figure~\ref{fig:detector_angle}). It shows an opposite nature for twisted field (see figure~\ref{fig:conc_detector separation}(b)).

Finally we fix the detector orientation $\alpha=\pi/2$ and analyze the density plot of  concurrence with varying energy gap and detector separation. In figure~\ref{fig:concurrence_density}, for untwisted fields in the single quotient space (c), the concurrence is lower than in Minkowski space near zero energy gap, thus making the difference $\mathcal{C}_{M_0}-\mathcal{C}_M$ negative in that region. However, this difference diminishes when two quotient spaces are symmetrically superposed in (a), indicating an increase in entanglement in the superposed spacetime. A similar trend is seen for positive energy gap, specifically for $
\sigma\Omega > 1$ and significantly large detector separation ($
a/\sigma > 0.8$), where the difference in concurrence is positive and even larger in the superposed spacetime. For twisted fields, on the other hand concurrence in single cylindrical space (f) is larger compared to Minkowski space near the origin of the energy gap and smaller in the region  $\sigma\Omega > 1$. These effects are even more pronounced in (d) for the superposed space compared to the single quotient space. This particularly suggests that entanglement is always greater in the superposed spacetime compared to Minkowski space and single quotient space. Intuitively, this behavior is somewhat expected. The action of the projector $\ket{L_i}\bra{L_i}$ in~\eqref{interaction Hamiltonian} on $\ket{\psi_i}$ induces mixing between various $\hat{\Phi}^{L_i}(x)\otimes \ket{L_i}$, in the final state $\ket{\psi_f}$, see~\eqref{App:final_state}, and this mixing introduces additional entanglement into the system.

Notably, in the superposed geometry, the entanglement region for the twisted field (d) is larger compared to other configurations (a,c,f). This can be understood by analyzing the contributing quantities \( X \) and \( P^E \) in the concurrence formula \( \mathcal{C}=\text{Max}~[0, 2(|X| - P^E)] \). The transition probability \( P^E \) for the twisted field near the origin of the energy gap is much smaller than for untwisted fields and even smaller in the superposed geometry compared to the single quotient space or regular Minkowski space (see the zoomed inset in figure~\ref{fig:probability-E}). The reasons for this are discussed in detail in the paragraph before section~\ref{sec:harvesting}. In contrast, the quantity \( X \) computed in~\eqref{x final} is simply the addition of \( X_M, X_{L_1} \), and \( X_{L_2} \) for \( \theta = \phi = \pi/4 \), and \( |X| \) in the symmetric superposed space is greater than in either Minkowski or the single quotient space. Thus, the combined effect of a larger \( |X| \) and smaller \( P^E \) in the superposed geometry for untwisted fields leads to a significantly higher concurrence, making the entanglement effect most pronounced in this case (refer to subfigure (d) in figure~\ref{fig:concurrence_density}).

\section{Conclusion and outlook}\label{sec:discussion}
We have explored \textit{entanglement harvesting} for a \textit{spacetime in quantum superposition} using two Unruh-DeWitt detectors coupled to a quantum scalar field. The spacetime background is modeled as a superposition of two quotient Minkowski spaces not related by diffeomorphisms, where the compactification introduces non-trivial topological effect and superposition introduces quantum mechanical effects.

While entanglement harvesting typically depends on the local properties of the quantum field vacuum, this study demonstrates that the \textit{global structure} of spacetime, specifically its superposed nature, can play a crucial role. The superposition of spacetime geometries introduces novel interference effects expressed through the Wightman functions $W^{L_1L_2}$ that modify the joint state of the detectors. The results reveal that superposed spacetime can significantly enhance the amount of entanglement harvested compared to just considering a single spacetime background~\cite{Martin-Martinez}. The interference between the modes of the quantum field, as dictated by the superposition parameters, introduces effects that amplify the concurrence function. Further the concurrence shows a clear dependence on the angles $(\theta, \phi)$ characterizing the superposed spacetime. The study further finds that when the control state of the spacetime is measured to be equal to initially prepared spacetime state, we get the maximum entanglement.

Our findings contribute to a better understanding how quantum gravitational phenomena, such as spacetime superposition, affect relativistic quantum information processing, such as entanglement harvesting. While we focused on the relatively simple setup involving superposed quotient Minkowski spaces, future work can explore superpositions of more complicated curved spacetimes or dynamical scenarios where the superposed geometries evolve over time. Such studies could shed light on the behavior of quantum fields and relativistic quantum information in more realistic spacetime models, potentially contributing to our understanding of quantum gravity. It may also serve as a starting point to explore experimental settings, such as spacetime superpositions induced by a superposition state of quantum particles with different masses. Such tabletop experiments may help to pave the way for deeper insights into the quantum nature of spacetime and its impact on quantum information processing such as \cite{Mart_n_Mart_nez_2014,howl2024quantumgravitycommunicationresource}.

\section{Acknowledgment}
AC expresses gratitude to J. Foo and A.R.H. Smith for valuable discussions. The research of AC and LH is supported by ‘The Quantum Information Structure of Spacetime’ Project (QISS), by grant $\#$63132 from the John Templeton Foundation. MZ acknowledges Knut and Alice Wallenberg foundation through a Wallenberg Academy Fellowship No. 2021.0119. LH acknowledges support by grant $\#$63132 from the John Templeton Foundation and an Australian Research Council Australian Discovery Early Career Researcher Award (DECRA) DE230100829 funded by the Australian Government. The opinions expressed in this publication are those of the authors and do not necessarily reflect the views of the respective funding organization.

\bibliography{biblio}

\clearpage

\onecolumngrid
\appendix
\section{Calculation of the components of $\rho_{AB}$}\label{app_density matrix}
 The unitary operators $U_n$s up to order of $\lambda^2$ in \eqref{finalstate} are given by
 \begin{align}
        U_0 &= \mathbf{I}\nonumber\\
        U_1 &= -\mi\int \,\,dt\,\,\Bigg[\Big(\frac{d\tau_A}{dt}\Big)\,H_A(\tau_A(t)) \otimes \mathbf{I}+\mathbf{I}\otimes \Big(\frac{d\tau_B}{dt}\Big)\,H_B(\tau_B(t))\Bigg]\nonumber\\
        U_2 &= - \hat{\mathcal{T}}\int \,dt\int\,dt'\Bigg[\frac{d\tau_A}{dt}\frac{d\tau_A}{dt'}H_A(\tau_A(t))H_A(\tau_A(t'))\otimes \mathbf{I}+ \mathbf{I}\otimes \frac{d\tau_B}{dt}\frac{d\tau_B}{dt'}H_B(\tau_B(t))H_B(\tau_B(t'))\nonumber\\
        &\qquad\qquad+\frac{d\tau_A}{dt}H_A(\tau_A(t))\otimes \frac{d\tau_B}{dt'} H_B(\tau_B(t'))+ \frac{d\tau_A}{dt'} H_A(\tau_A(t'))\otimes \frac{d\tau_B}{dt}H_B(\tau_B(t))\Bigg]
    \end{align}
     The associated $\ket{\psi_f^{(n)}}$s up to order of $\lambda^2$ are given by
       \begin{align}
    \ket{\psi_f^{(0)}}&=\ket{\psi_i}\nonumber\\
    \ket{\psi_f^{(1)}}&=-\mi\int_{-\infty}^{\infty} dt\Bigg[\ket{1,0} \otimes \eta_A(t) \e^{\mi\, \Omega_A\tau_A(t)}\Big\{\cos\theta~\hat{\Phi}^{L_1}(x_A)\ket{0}_F\otimes \ket{L_1} +\sin\theta~\hat{\Phi}^{L_2}(x_A)\ket{0}_F\otimes \ket{L_2}\Big\}+
\ket{0,1} \otimes (A\leftrightarrow B)\Bigg],\nonumber\\
        \ket{\psi_f^{(2)}}&= -\int_{-\infty}^{\infty} dt\int_{-\infty}^t dt' \Bigg[\ket{0,0} \otimes\,\Bigg\{\eta_A(t)\eta_A(t') \e^{-\mi\, \Omega_A (\tau_A(t)-\tau_A(t'))}\Big(\cos\theta\,\hat{\Phi}^{L_1}(x_A)\hat{\Phi}^{L_1}(x'_A)\ket{0}_F\otimes \ket{L_1}\nonumber\\
        &+\sin\theta\,\hat{\Phi}^{L_2}(x_A)\hat{\Phi}^{L_2}(x'_A)\ket{0}_F\otimes \ket{L_2}\Big) +(A \leftrightarrow B)\Bigg\}+2\ket{1,1} \otimes\eta_A(t)\eta_B(t') \e^{\mi\, [\Omega_A(\tau_A(t))+\Omega_B(\tau_B(t'))]}\nonumber\\
        & \Big(\cos\theta\,\hat{\Phi}^{L_1}(x_A)\hat{\Phi}^{L_1}(x'_B) \ket{0}_F \otimes \ket{L_1} +\sin\theta\,\hat{\Phi}^{L_2}(x_A)\hat{\Phi}^{L_2}(x'_B) \ket{0}_F \otimes \ket{L_2}\Big)\Bigg],\label{App:final_state}
    \end{align}
    where we used $\eta_A(t)=\chi_A(\tau_A)\frac{d\tau_A}{dt}$ and $\hat{\Phi}^{L_i}(x_D)\equiv\hat{\Phi}^{L_i}(x_D(\tau_D(t))),~\hat{\Phi}^{L_i}(x'_D)\equiv\hat{\Phi}^{L_i}(x_D(\tau_D(t')))$.
    Now tracing out over all possible fields and taking measurement in the final spacetime control state $\ket{s_f}$ we can find out the components of the joint detector state given by
    \begin{align}
    &\Tr_{\Phi}[\braket{s_f|\psi_f^0}\braket{\psi_f^0|s_f}]=(a+b)^2\ket{0,0}\bra{0,0}\,,\\
      &\Tr_{\Phi} [\langle s_f|\psi_f^1\rangle\langle\psi_f^1|s_f\rangle]\nonumber\\
      &= \ket{1,0}\bra{1,0}\int dt\,dt' \eta_A(t)\eta_A(t')\e^{\mi\, \Omega_A(\tau_A(t)-\tau_A(t'))}\Big[a^2~W^{L_1}(x'_A,x_A)+b^2~W^{L_2}(x'_A,x_A)+2ab~W^{L_1L_2}(x'_A,x_A)\Big]\nonumber\\
      &+ \ket{0,1}\bra{0,1} \int dt\,dt' \eta_B(t)\eta_B(t')\e^{\mi\, \Omega_B(\tau_B(t)-\tau_B(t'))}\Big[a^2~W^{L_1}(x'_B,x_B)+b^2~W^{L_2}(x'_B,x_B)+2ab~W^{L_1L_2}(x'_B,x_B)\Big]\nonumber\\
      &+\ket{1,0}\bra{0,1}\int dt\,dt' \eta_A(t)\eta_B(t')\e^{\mi\, [\Omega_B\tau_B(t)-\Omega_A\tau_A(t')]}\Big[a^2~W^{L_1}(x'_B,x_A)+b^2~W^{L_2}(x'_B,x_A)+ 2ab~W^{L_1L_2}(x'_B,x_A)\Big]\nonumber\\
      &+\ket{0,1}\bra{1,0} \int dt\,dt' \eta_B(t)\eta_A(t')\e^{\mi\, [\Omega_A\tau_A(t)-\Omega_B\tau_B(t')]}\Big[a^2~W^{L_1}(x'_A,x_B)+b^2~W^{L_2}(x'_A,x_B)+ 2ab~W^{L_1L_2}(x'_A,x_B)\Big]\\
&\Tr_{\Phi}[\langle s_f|\psi_f^2\rangle\langle\psi_f^0| s_f\rangle]\nonumber\\
&= -|0,0\rangle\langle 0,0|(a^2+ab)\Big[\int dt\,dt'\eta_A(t)\eta_A(t')\e^{-\mi\, \Omega_A(\tau_A(t)-\tau_A(t'))}\Big[W^{L_1}(x_A(t),x_A(t'))+W^{L_2}(x_A(t),x_A(t'))\Big]\nonumber\\
&+(b^2+ab)(A\leftrightarrow B)\Big]-|1,1\rangle\langle 0,0|\Bigg[(a^2+ab)\int dt\,dt'\eta_A(t)\eta_B(t')\e^{\mi\, [\Omega_A\tau_A(t)+\Omega_B\tau_B(t')]}\Big[W^{L_1}(x_A(t),x_B(t'))\nonumber\\
&\hspace{250pt}+W^{L_2}(x_A(t),x_B(t'))\Big]+(b^2+ab)(A\leftrightarrow B)\Big]
\end{align}
where $a=\cos\theta\cos\phi,\, b=\sin\theta\sin\phi$. Likewise one can also calculate $\Tr[\braket{s_f|\psi_f^0}\braket{\psi_f^2|s_f}]_{\Phi}$. Then accumulating the coefficients of the bases $\ket{0,0}\bra{0,0},\ket{0,1}\bra{0,1}, \ket{1,0}\bra{1,0}$ and $\ket{1,1}\bra{1,1}$  one can now easily construct the density matrix $\rho_{AB}$ as \eqref{density matrix} where the components are given by~\eqref{pground}-\eqref{x term} and calculated as below.
\subsection{Evaluation of transition probabilities}
For the following calculations we consider two identical detectors with $\eta_A(t)=\eta_B(t)=\e^{-\frac{t^2}{2\sigma^2}}$ and $ \Omega_A=\Omega_B=\Omega$. With these considerations the transition probability of a detector in a single cylindrical spacetime is calculated as
\begin{align} 
  & P^{L_i}
   =\int dt\,dt' \eta(t)\eta(t')\e^{-\mi\, \Omega(t-t')}~W^{L_i}(x,x')\nonumber\\
   &=\int\,dt\,dt'\, \e^{-\frac{t^2+t'^2}{2\sigma^2}}\e^{-\mi\Omega(t-t')} \frac{1}{\mathcal{N}}\sum_{m,n=-\infty}^{\infty} \gamma^n\gamma^m \langle 0|\hat{\phi}(J_{L_i}^nx)\hat{\phi}(J_{L_i}^mx')|0\rangle,\qquad \gamma=\pm 1 \textrm{for untwisted and twisted fields}\nonumber\\
   &=\frac{1}{\mathcal{N}}\int du\,ds \e^{-\frac{u^2}{4\sigma^2}}\e^{-\frac{s^2}{4\sigma^2}}\e^{-\mi\, \Omega s}\sum_{n,m} \gamma^n\gamma^m W_M(J^n_{0_{L_i}}x,J^m_{0_{L_i}}x'),\label{A1}
\end{align}
where we have used $u=t+t'$ and $s=t-t'$. The Wightman function in Minkowski space is well known and given by $W_M(x,x')=\frac{\sgn(s)\delta(s^2)}{4\pi\mi}-\frac{1}{4\pi^2s^2}$.
One can manipulate the term in the summation along with the normalization outside in~\eqref{A1} as
\begin{align}
\frac{1}{\mathcal{N}}\sum_{n,m} \gamma^n\gamma^m W_M(J^n_{0_{L_i}}x,J^m_{0_{L_i}}x')&=
   \frac{1}{\mathcal{N}} \sum_{n, m} \gamma^n (\gamma^n \gamma^m) W_M(J_0^n x, J_{0_{L_i}}^n J_{0_{L_i}}^m x') \nonumber\\
    &= \frac{1}{\mathcal{N}} \sum_{n, m} \gamma^{2n} \gamma^m W_M(x,J_{0_{L_i}}^m x'),\,\, \textrm{using translation invariance}\nonumber\\
    &= \sum_m \gamma^m W_M(x, J_{0_{L_i}}^m x')\nonumber\\
    &=W_M(x,x')+\sum_{m\ne 0}\gamma^m W_M(x,J_{0_{L_i}}x').
\end{align}
Thus we can further simplify \eqref{A1} to find the individual transition probability in single quotient space as
\begin{align}
   P^{L_i}&=P^M+\sqrt{\pi}\sigma\sum_{m\ne 0} \int_{-\infty}^{\infty} ds\gamma^m\,\e^{-\frac{s^2}{4\sigma^2}}\e^{-\mi\, \Omega s}\Big[\frac{\sgn(s)\delta(s^2-m^2L_i^2)}{4\pi\mi}-\frac{1}{4\pi^2(s^2-m^2L_i^2)}\Big] \nonumber\\
   &=P^M+\frac{\sigma}{2\sqrt{\pi}}\sum_{m=1}^{\infty}\gamma^m\frac{\e^{-m^2L_i^2/4\sigma^2}}{mL_i}\Big[\im\Big(\e^{\mi mL_i\Omega }\,\erf \Big(\frac{\mi mL_i}{2\sigma}+\sigma\Omega\Big)\Big)-\sin(\Omega mL_i)\Big]\label{trans prob final}
\end{align}
where $P^M=\frac{1}{4\pi}\Big[\e^{-\sigma^2\Omega^2}-\sqrt{\pi}\sigma\Omega\,\erfc(\sigma\Omega)\Big]$ (for detail calculation refert to~\cite{Foo_super_Minkowski}). The interference term $P^{L_1L_2}$ in transition probability in the superposed spacetime state is given by
\small
\begin{align}
    &P^{L_1L_2} =\frac{\sqrt{\pi}\sigma}{\mathcal{N}}\sum_{n,m}\int ds\gamma^{n+m}\,\e^{-\frac{s^2}{4\sigma^2}}\e^{-\mi\, \Omega s}\,\Big\langle 0\Big|\,\hat{\phi}(J^n_{0_{L_1}}x)\,\hat{\phi}(J^m_{0_{L_2}}x')\,\Big|0\Big\rangle\nonumber\\
    &=\frac{1}{\mathcal{N}}\sum_{L_1n=L_2m}\gamma^{n+m}P^M + \frac{\sqrt{\pi}\sigma}{\mathcal{N}}\sum_{L_1n\ne L_2m}\int_{-\infty}^{\infty} ds\,\gamma^{n+m}\e^{-\frac{s^2}{4\sigma^2}}\e^{-\mi\, \Omega s}\Big[\frac{\sgn(s)\delta(s^2-(L_1n-L_2m)^2)}{4\pi\mi}-\frac{1}{4\pi^2(s^2-(L_1n-L_2m)^2)}\Big]\nonumber\\
    &=\frac{1}{\mathcal{N}}\sum_{L_1n=L_2m}\gamma^{n+m}P^M+\frac{\sigma}{2\sqrt{\pi}\mathcal{N}}\sum_{L_1n\ne L_2m}\gamma^{n+m}\frac{\e^{-(L_1n-L_2m)^2/4\sigma^2}}{(L_1n-L_2m)}\Big[\im\Big(\e^{\mi\, (L_1n-L_2m)\Omega }\,\erf\Big(\frac{\mi(L_1n-L_2m)}{2\sigma}+\sigma\Omega\Big)\Big)\nonumber\\&\hspace{330pt}-\sin(\Omega (L_1n-L_2m))\Big]\,\label{A9}.
\end{align}
\normalsize
Here one has to be cautious while dealing with the Wightman function in superposed space and handling of the normalization constant. Note that for $nL_1=mL_2$ we can single out the Minkowskian Wightman function, however we have to take into account all the possibilities of $nL_1=mL_2$ for $n,m=-\infty$ to $\infty$. Technically that should generate infinite number of terms giving diverging result, however the normalization in the denominator being itself a diverging quantity regularizes the sum giving a finite value. The same reasoning applies for the second term also when $nL_1\ne mL_2$ for various combinations of $n,m$. However, the regularization in the first and second terms of \eqref{A9} proceeds differently and one cannot simply cancel $\mathcal{N}$ in the denominator with the infinite number of terms in the numerator (see figure~\ref{Fig:lattice} for a pictorial explanation). In our numerical computation we have used a finite lattice size for $n=\pm 20, m=\pm 20$ to produce the figures~\ref{fig:probability-E} to figure~\ref{fig:concurrence_density}, as including additional terms does not significantly affect the plots. This is because the contribution from larger $n, m$ values in the image sums decreases rapidly as $n, m$ increases, following the behavior $\frac{e^{-n^2}}{n}$, as evident from~\eqref{pE final} and~\eqref{x final}. 
\begin{center}
\begin{figure}
\begin{tikzpicture}[scale=0.3, line cap=round, line join=round]
  Draw the x and y axes
    \draw[<->] (-10.5,0) -- (10.5,0) node[right] {$m$};
    \draw[<->] (0,-10.5) -- (0,10.5) node[above] {$n$};

    \foreach \x in {-10,...,10} {
        \foreach \y in {-10,...,10} {
            \filldraw (\x,\y) circle (1.5pt);
        }
    }
       \foreach \x in {-10,-8,-6,-4,-2,0,2,4,6,8,10} {
        \pgfmathsetmacro{\y}{\x/2}
        \filldraw[red] (\x,\y) circle (4.5pt);
    }
\end{tikzpicture}
\caption{Above is an example of \(20 \times 20\) lattice, where points satisfying \(nL_1 = mL_2\) are indicated in red for \(L_1 = 1\) and \(L_2 = 0.5.\) For this lattice, the normalization is \(\mathcal{N} = 21\), with \(nL_1 = mL_2\) occurring 11 times and \(nL_1 \neq mL_2\) occurring 389 times. Although all these numbers tend to infinity as \(n,m \to \infty\), clearly they do not scale in the same way. This  is the reason we must carefully count them when numerically evaluating \eqref{A9} and \eqref{A21}.
}\label{Fig:lattice}
\end{figure}
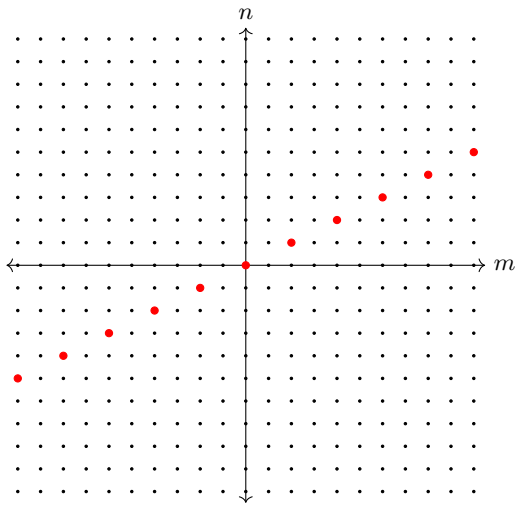
\end{center} 
Finally, the total probability of transition~\eqref{total transprob} in a superposed space is given by
\begin{align}
  & P^E=(a^2+b^2)P^M+\frac{a^2\sigma}{2\sqrt{\pi}}\sum_{m=1}^{\infty}\frac{\e^{-m^2L_1^2/4\sigma^2}}{mL_1}\Big[\im\Big(\e^{\mi mL_1\Omega }\,\erf \Big(\frac{\mi mL_1}{2\sigma}+\sigma\Omega\Big)\Big)-\sin(\Omega mL_1)\Big]\nonumber\\
   &+\frac{b^2\sigma}{2\sqrt{\pi}}\sum_{m=1}^{\infty}\frac{\e^{-m^2L_2^2/4\sigma^2}}{mL_2}\Big[\im\Big(\e^{\mi mL_2\Omega }\,\erf \Big(\frac{\mi mL_2}{2\sigma}+\sigma\Omega\Big)\Big)-\sin(\Omega mL_2)\Big]
  +\frac{2ab}{\mathcal{N}}\sum_{L_1n=L_2m}P^M\nonumber\\&+\frac{2ab\sigma}{2\sqrt{\pi}\mathcal{N}}\sum_{L_1n\ne L_2m}\frac{\e^{-(L_1n-L_2m)^2/4\sigma^2}}{(L_1n-L_2m)}\Big[\im\Big(\e^{\mi\, (L_1n-L_2m)\Omega }\,\erf\Big(\frac{\mi(L_1n-L_2m)}{2\sigma}+\sigma\Omega\Big)\Big)-\sin(\Omega (L_1n-L_2m))\Big]  \label{pE final}
\end{align}
with $a=\cos\theta\cos\phi,\,\,b=\sin\theta\sin\phi$.

\subsection{Evaluation of matrix component X}
Let us start by deriving the individual terms of $X=(a^2+ab)X_{L_1}+(b^2+ab)X_{L_2}$ in~\eqref{x term} given by 
\begin{equation}
 X_{L_i}= \int dt\,dt'\e^{-\frac{(t+t')^2}{2\sigma^2}}\e^{-\mi\Omega(t+t')}~~W^{L_i}(x'_B,x_A),\quad i=1,2 ,
\end{equation}
where the Wightman function $W^{L_i}(x'_B,x_A)$  ~\eqref{wightmansingle} is given by
\begin{equation}
W^{L_i}(x'_B,x_A)=\sum_{m = -\infty}^\infty \gamma^m\left[
\frac{1}{4\pi \mi} \, \mathrm{sgn}(t' - t) \, \delta\left[(t' - t)^2 - L_i(m)^2\right]
- \frac{1}{4\pi^2 \left[(t' - t)^2 - L_i(m)^2\right]}
\right]\,,\label{x sum}
\end{equation}
where $L_i(m)=\sqrt{a^2+m^2L_i^2+2 a m L_i \sin\alpha}$ (refer to figure~\ref{fig:detector_angle}) is the distance between two detectors situated at $J_{0_{L_i}}^n x_A$ and $J_{0_{L_i}}^m x'_B$ in quotient Minkowski space with compactification length $L_i$ and $a^2=(x_A-x_B)^2+(y_A-y_B)^2+(z_A-z_B)^2$ is the square of the spatial distance between two detectors in regular Minkowski space. 
We begin by evaluating the $m=0$ terms in~\eqref{x sum} to get  the corresponding term for Minkowski space given by
\begin{equation}
    X_M=-\lambda^2\int dt\,dt'\eta(t)\eta(t')\e^{-\mi \Omega (t+ t')} \Bigg[\frac{\sgn(t'-t)\delta[(t'-t)^2-a^2]}{4\pi\mi}-\frac{1}{4\pi^2[(t'-t)^2-a^2]}\Bigg]\,.\label{x_minkowski}
\end{equation}
Changing the integration variables to \( u' = t' + t \) and \( u = t' - t \), the quantity \( X \) simplifies to:
\begin{align}
X_M&= -\lambda^2 \int_{-\infty}^\infty du' \, e^{-u'^2/4\sigma^2} e^{-i\Omega u'} \int_{-\infty}^0 du \, e^{-u^2/4\sigma^2} 
\left( 
\frac{1}{4\pi \mi} \, \mathrm{sgn}(u)\delta(u^2 - a^2) - \frac{1}{4\pi^2(u^2 - a^2)} 
\right)\nonumber\\
&= -\lambda^2 \int_{-\infty}^\infty du' \, e^{-u'^2/4\sigma^2} e^{-\mi\Omega u'} \int_0^\infty du \, e^{-u^2/4\sigma^2} 
\left( 
\frac{1}{4\pi \mi} \, \mathrm{sgn}(-u)\delta(u^2 - a^2) - \frac{1}{4\pi^2(u^2 - a^2)} 
\right)\nonumber\\
&= 2\sqrt{\pi} \lambda^2 \sigma e^{-\sigma^2 \Omega^2} \int_0^\infty du \, e^{-u^2/4\sigma^2} 
\left( 
\frac{1}{4\pi \mi} \, \mathrm{sgn}(-u)\delta(u^2 - a^2) + \frac{1}{4\pi^2(u^2 - a^2)} 
\right)\nonumber\\
&= 2\sqrt{\pi} \lambda^2 \sigma e^{-\sigma^2 \Omega^2} \int_0^\infty du \, e^{-u^2/4\sigma^2} 
\left( 
-\frac{1}{4\pi \mi} \frac{\delta[u - a]}{2a} + \frac{1}{4\pi^2(u^2 - a^2)} 
\right)\nonumber\\
&= 2\sqrt{\pi} \lambda^2 \sigma e^{-\sigma^2 \Omega^2} 
\left( 
-\frac{1}{4\pi \mi} \frac{e^{-a^2/4\sigma^2}}{2a} + \frac{1}{4\pi} \frac{\mi}{2a} e^{-a^2/4\sigma^2} \mathrm{erf}\left(\frac{\mi a}{2\sigma}\right)
\right)\nonumber\\
& = \mi \frac{\lambda^2}{4\sqrt{\pi}} \frac{\sigma}{a} e^{-\sigma^2 \Omega^2 - a^2/4\sigma^2} 
\left( 
\mathrm{erf}\left(\frac{\mi a}{2\sigma}\right) +1
\right).
\end{align}
where the second integration was performed with Mathematica. Upon comparing~\eqref{x sum} and~\eqref{x_minkowski} it is clear that each term in the rest of the sum in~\eqref{x sum} for $ m\ne 0$ is equivalent to the Minkowski term $X_M$~\eqref{x_minkowski} if we just replace $a$ with $L_i(m)$ giving us the identical result as
\begin{align}
\mi\frac{\lambda^2}{4\sqrt{\pi}}\sum_{m\ne 0} \gamma^m\frac{\sigma}{L_i(m)}\e^{-\sigma^2\Omega^2-\frac{L_i(m)^2}{4\sigma^2}}\Big[\erf\Big(\mi\frac{L_i(m)^2}{2\sigma}\Big)-1\Big].
    \end{align}
    So the total expression of $X$ in~\eqref{x term} boils down to
    \begin{align}
        X=(a+b)^2X_M&+\mi(a^2+ab) \frac{\lambda^2}{4\sqrt{\pi}}\sum_{m\ne 0} \gamma^m\frac{\sigma}{L_1(m)}\e^{-\sigma^2\Omega^2-\frac{L_1(m)^2}{4\sigma^2}}\Big[\erf\Big(\mi\frac{L_1(m)^2}{2\sigma}\Big)-1\Big]\nonumber\\
        &+\mi(b^2+ab) \frac{\lambda^2}{4\sqrt{\pi}}\sum_{m\ne 0}\gamma^m \frac{\sigma}{L_2(m)}\e^{-\sigma^2\Omega^2-\frac{L_2(m)^2}{4\sigma^2}}\Big[\erf\Big(\mi\frac{L_2(m)^2}{2\sigma}\Big)-1\Big].\label{x final}
    \end{align}

\subsection{Expression of matrix component C}
Although we have not explicitly used this component in computation of entanglement \ie concurrence, for the sake of completeness we briefly show the computation. We start by evaluating the individual terms of $C=a^2C_{L_1}+b^2C_{L_2}+2abC_{L_1L_2}$ in~\eqref{c term} with
\begin{align}
C_{L_i}&=\lambda^2 \int dt' dt \, e^{-(t^2 + t'^2)/4\sigma^2} e^{-i\Omega(t - t')} W^{L_i}(x_A,x'_B)\,,\label{c1 term}\\
C_{L_1L_2}&=\lambda^2 \int dt' dt \, e^{-(t^2 + t'^2)/4\sigma^2} e^{-i\Omega(t - t')} W^{L_1L_2}(x_A,x'_B)\,,\label{c3 term}
\end{align} 
where $W^{L_i}(x_A, x'_B)$ and $W^{L_1L_2}(x_A,x'_B)$ are defined in~\eqref{wightmansingle} and~\eqref{wightmandouble}. To calculate \eqref{c1 term} we again use \eqref{x sum} for $W^{L_i}(x_A, x'_B)$ and first separate the terms $m=0$ in sum~\eqref{x sum} to calculate the Minkowski contribution to $C$ as 
\begin{align}
C_M&=\lambda^2 \int du' \, du \, e^{-u'^2/4\sigma^2} e^{-u^2/4\sigma^2} e^{-\mi\Omega u}\left[
\frac{1}{4\pi \mi} \mathrm{sgn}(u) \delta(u^2 - a^2) - \frac{1}{4\pi^2 (u^2 - a^2)}
\right]\quad\nonumber\\
&=\sqrt{\pi} \lambda^2 \sigma e^{-\sigma^2 \Omega^2} 
\int du \, e^{-u^2/4\sigma^2} 
\left[
\frac{1}{4\pi \mi} \mathrm{sgn}(u) \frac{\delta(u + a) + \delta(u - a)}{2a} 
- \frac{1}{4\pi^2 (u^2 - a^2)}
\right]\nonumber\\
&= -\frac{\lambda^2}{4\sqrt{\pi}} \frac{e^{-\frac{a^2}{4\sigma^2}}}{a} \sin(\Omega a) + \frac{1}{\pi} \int du \, \frac{e^{-u^2/4\sigma^2} e^{-\mi\Omega u}}{u^2 - a^2}\nonumber\\
&= \frac{\lambda^2}{4\sqrt{\pi}} \frac{\sigma}{a} e^{-\frac{a^2}{4\sigma^2}} 
\left[
\mathrm{Im} \left( e^{\mi a\Omega} \mathrm{erf} \left(\frac{a}{2\sigma} + \sigma \Omega \right) \right) - \sin(\Omega a)
\right]. \label{c minkowski}
\end{align}
where we have in the first line $u' = t + t'$ and $,u = t-t'$. To calculate the terms with $m\ne 0$ in the sum~\eqref{x sum}, we replace $a$ in~\eqref{c minkowski} with $L_i(m)$ to find~\eqref{c1 term} as 
\begin{equation}
C_{L_i}= C_M+ \frac{\lambda^2}{4\sqrt{\pi}}\sum_{m\ne 0} \gamma^m  \frac{\sigma}{L_i(m)} \e^{-\frac{L_i(m)^2}{4\sigma^2}}\Bigg(\im\Big[\e^{\mi  L_i(m)\Omega} \erf\Big(\mi\frac{L_i(m)}{2\sigma}+\sigma\Omega\Big)\Big]-\sin \Big(\Omega L_i(m)\Big)\Bigg).
\end{equation}
To evaluate ~\eqref{c3 term} we make use of the Wightman function in~\eqref{wightmandouble} and replace $L_i(n,m)$ in~\eqref{x sum} with $L_{12}(n,m)=\sqrt{a^2+(nL_1-mL_2)^2+2 a (nL_1-mL_2)\sin\alpha}$ and evaluate it as
\small
\begin{equation}
    C_{L_1L_2}= \frac{1}{\mathcal{N}}\sum_{nL_1=mL_2} \gamma^{n+m}C_M+\frac{\lambda^2}{4\sqrt{\pi}\mathcal{N}}\sum_{nL_1\ne mL_2} \gamma^{n+m}  \frac{\sigma \e^{-\frac{L_{12}(n,m)^2}{4\sigma^2}}}{L_{12}(n,m)} \Bigg(\im\Big[\e^{\mi  L_{12}(n,m)\Omega} \erf(\mi\frac{L_{12}(n,m)}{2\sigma}+\sigma\Omega)\Big]-\sin \Big(\Omega L_{12}(n,m)\Big)\Bigg).\label{A21}
\end{equation}
\normalsize
So finally the total matrix component $C$ is given by
\begin{align}
    C&=(a^2+b^2)C_M+\frac{a^2\lambda^2}{4\sqrt{\pi}}\sum_{m\ne 0}\gamma^m   \frac{\sigma}{L_1(m)} \e^{-\frac{L_1(m)^2}{4\sigma^2}}\Bigg(\im\Big[\e^{\mi  L_1(m)\Omega} \erf\Big(\mi\frac{L_1(m)}{2\sigma}+\sigma\Omega\Big)\Big]-\sin \Big(\Omega L_1(m)\Big)\Bigg)\nonumber\\
    &+\frac{b^2\lambda^2}{4\sqrt{\pi}}\sum_{m\ne 0} \gamma^m  \frac{\sigma}{L_2(m)} \e^{-\frac{L_2(m)^2}{4\sigma^2}}\Bigg(\im\Big[\e^{\mi  L_2(m)\Omega} \erf\Big(\mi\frac{L_2(m)}{2\sigma}+\sigma\Omega\Big)\Big]-\sin \Big(\Omega L_2(m)\Big)\Bigg)+
  \frac{2ab}{\mathcal{N}}\sum_{nL_1=mL_2} \gamma^{n+m}C_M\nonumber\\&+\frac{2ab\lambda^2}{4\sqrt{\pi}\mathcal{N}}\sum_{nL_1\ne mL_2}  \gamma^{n+m} \frac{\sigma \e^{-\frac{L_{12}(n,m)^2}{4\sigma^2}}}{L_{12}(n,m)} \Bigg(\im\Big[\e^{\mi  L_{12}(n,m)\Omega} \erf(\mi\frac{L_{12}(n,m)}{2\sigma}+\sigma\Omega)\Big]-\sin \Big(\Omega L_{12}(n,m)\Big)\Bigg).\label{c final}
\end{align}
\section{Normalization of density matrix $\rho_{AB}$}\label{normalization}
To obtain a normalized density matrix, we simply divide the components of $\rho_{AB}$ \eqref{density matrix} by its trace. In particular  for $\ket{s_i}=\ket{s_f}$ we take $\theta=\phi$ and  the corresponding trace of the un-normalised density matrix is given (up to $\mathcal{O}(\lambda^2)$) as
\begin{align}
&\Tr(\rho_{AB})=N
\simeq\,1-\lambda^2 ab\Big(P_A^{L_1}+P_A^{L_2}+P_B^{L_1}+P_B^{L_2}-2P_A^{L_1L_2}-2P_B^{L_1L_2}\Big),
\end{align}
where $a=\cos^2\theta,\,b=\sin^2\theta$.
The entries of the normalized density matrix $\tilde{\rho}_{AB}$ up to order $\lambda^2$ are now given by
\begin{align}
   & \tilde{P}^G= P^G N^{-1}=1-\lambda^2\Big[a^2
   (P_A^{L_1}+P_B^{L_1})+b^2(P_A^{L_2}+P_B^{L_2})+2ab(P_A^{L_1L_2}+P_B^{L_1L_2})\Big]\,,\nonumber\\
   & \tilde{P}_D^{E}= P_D^{E}N^{-1}=P_D^E,\,\tilde{C}= C N^{-1}=C,\,\,
      \tilde{X}=XN^{-1}=X\,.
\end{align}
However, for further use we shall omit the tilde and use the original notations in~\eqref{xstate}.

\end{document}